%% file: main.tex
\documentclass[year=25,pdfa]{fmcad}
\IEEEoverridecommandlockouts

\def\BibTeX{{\rm B\kern-.05em{\sc i\kern-.025em b}\kern-.08em
    T\kern-.1667em\lower.7ex\hbox{E}\kern-.125emX}}

\input{init}

\begin{document}

\title{\polyverl: A Compositional Approach for \\ 
Polyglot System Modeling and Verification}



\author{
    \IEEEauthorblockN{
    Pei-Wei Chen\IEEEauthorrefmark{1}\orcid{0000-0002-8671-3065}, 
    Shaokai Lin\IEEEauthorrefmark{1}\orcid{0000-0001-6885-5572}, 
    Adwait Godbole\IEEEauthorrefmark{1}\orcid{0000-0001-7704-304X}, 
    Ramneet Singh\IEEEauthorrefmark{2}\orcid{0009-0001-3204-7562},
    \\
    Elizabeth Polgreen\IEEEauthorrefmark{3}\orcid{0000-0001-9032-7661}, 
    Edward A. Lee\IEEEauthorrefmark{1}\orcid{0000-0002-5663-0584}, 
    Sanjit A. Seshia\IEEEauthorrefmark{1}\orcid{0000-0001-6190-8707}
    }
    \IEEEauthorblockA{\IEEEauthorrefmark{1}University of California, Berkeley, USA, \{pwchen, shaokai, adwait, eal, sseshia\}@berkeley.edu}
    \IEEEauthorblockA{\IEEEauthorrefmark{2}Indian Institute of Technology, Delhi, India, ramneet2001@gmail.com}
    \IEEEauthorblockA{\IEEEauthorrefmark{3}University of Edinburgh, UK, elizabeth.polgreen@ed.ac.uk}
}


\maketitle

\input{abstract}

\input{introduction}

\input{background}

\input{problem_statement}

\input{4-verification-through-contracts}

\input{approach}

\input{implementation}

\input{case_study}

\input{evaluation}

\input{related_work}

\input{conclusion}

\input{acknowledgement}

\bibliographystyle{IEEEtran}
\bibliography{refs}

\clearpage
\appendix
\input{appendix}

\end{document}

%% file: init.tex
\usepackage{times}
\usepackage{cite}
\usepackage{amsmath,amssymb,amsfonts,amsthm}
\usepackage{graphicx}
\usepackage{textcomp}

\usepackage[T1]{fontenc}
\usepackage{graphicx}
\usepackage{color}
\usepackage{xspace}
\usepackage{stmaryrd}
\usepackage{algorithm}
\usepackage{algpseudocode}
\usepackage{todonotes}
\usepackage{booktabs}
\usepackage{listings}
\usepackage{subcaption}

\usepackage{hyperref}

\usepackage{wrapfig}

\usepackage{multicol}

\usepackage[switch]{lineno}

\definecolor{codegreen}{rgb}{0,0.6,0}
\definecolor{codegray}{rgb}{0.5,0.5,0.5}
\definecolor{codepurple}{rgb}{0.58,0,0.82}
\definecolor{backcolour}{rgb}{0.95,0.95,0.92}

\lstdefinestyle{mystyle}{
    backgroundcolor=\color{backcolour},   
    commentstyle=\color{codegreen},
    keywordstyle=\color{magenta},
    numberstyle=\tiny\color{codegray},
    stringstyle=\color{codepurple},
    basicstyle=\ttfamily\footnotesize,
    breakatwhitespace=false,         
    breaklines=true,                 
    captionpos=b,                    
    keepspaces=true,                 
    numbers=left,                    
    numbersep=5pt,                  
    showspaces=false,                
    showstringspaces=false,
    showtabs=false,                  
    tabsize=2
}
\newtheorem{definition}{Definition}
\newtheorem{theorem}{Theorem}
\newtheorem{problem_statement}{Problem Statement}

\newcommand{\polyverl}{\textsc{PolyVer}\xspace}

\def\tl{{\mathcal{L}}}
\def\intp#1{{\llbracket #1 \rrbracket}}
\def\v#1{#1}
\def\il{{\mathcal{L}^*}}

\def\model{{\mathcal{M}}}

\newcommand{\poseg}{X}
\newcommand{\negeg}{\overline{X}}

\newcommand{\prquery}{{\color{orange} \texttt{ProcessQuery}\xspace}}

\newcommand{\prwaited}{{\color{blue} \texttt{Waited}}\xspace}

\newcommand{\compile}{\mathsf{comp}}
\newcommand{\synthoracle}{\mathsf{synth}}
\newcommand{\verifyoracle}{\mathsf{verif}}
\newcommand{\modelcheck}{\mathsf{mc}}
\newtheorem{tcb}{Assumption}

\newcounter{myctr}

\newenvironment{myitemize}{\begin{list}{$\bullet$}
{\setlength{\topsep}{1mm}\setlength{\itemsep}{0.25mm}
\setlength{\parsep}{0.1mm}
\setlength{\itemindent}{0mm}\setlength{\partopsep}{0mm}
\setlength{\labelwidth}{15mm}
\setlength{\leftmargin}{4mm}}}{\end{list}}

\algnewcommand{\Modify}[1]{\item[\textbf{Modify:}] #1}

\definecolor{lfkeywordcolor1}{RGB}{0,0,180} 
\definecolor{lfkeywordcolor2}{rgb}{0.58,0,0.82} 
\definecolor{lfstringcolor}{RGB}{163,21,21} 
\definecolor{lfcommentcolor}{RGB}{0,128,0} 

\lstdefinelanguage{lfLang}{
    morekeywords=[1]{reactor,input,output,state,logical,action,bool,u32,uint32_t,true,false,main,startup},       
    morekeywords=[2]{reaction,if,else,new},       
    morekeywords=[3]{set,lf_set,saturating_add,get,schedule,unwrap,rand},       
    sensitive=true,                   
    morecomment=[l]{//},              
    morecomment=[s]{/*}{*/},          
    morestring=[b]"                   
}

\lstdefinestyle{lfStyle}{
    language=lfLang,
    basicstyle=\ttfamily\scriptsize,
    keywordstyle=[1]\color{lfkeywordcolor1},
    keywordstyle=[2]\color{lfkeywordcolor2},
    keywordstyle=[3]\color{brown},
    commentstyle=\color{lfcommentcolor}\itshape,
    stringstyle=\color{lfstringcolor},
    columns=fullflexible,
    tabsize=2,
    showstringspaces=false
}

%% file: abstract.tex
\begin{abstract}
Many software systems are {\em polyglot}; that is, they comprise programs implemented in a combination of programming languages.
Program verifiers, however, tend to be customized for individual languages.
Verification by compiling to a common encoding requires supporting full language syntax and semantics which is prohibitive for modern languages.
%
We present \polyverl, an alternative compositional approach to polyglot verification that bootstraps off-the-shelf language-specific verifiers with abstraction and synthesis.
%
%

\polyverl uses contracts written in an intermediate language to abstract individual procedures in the system.
Our verification approach uses language-specific verifiers (e.g., for C or Rust) to validate these contracts and the UCLID5 model checker for compositionally verifying a temporal property on the overall system using the contracts.
The intermediate language sidesteps the need for compiling implementation languages to a common encoding, a key obstacle with polyglot verification.
Finally, \polyverl automates the generation of contracts using synthesis oracles such as large-language-models (LLMs).
Overall \polyverl performs contract synthesis and verification in a counterexample-guided abstraction refinement and inductive synthesis (CEGIS-CEGAR) loop to verify the system-level property.
%
%
We use \polyverl to verify programs in the Lingua Franca polyglot language. 
We are able to verify systems with C and Rust procedures, as well as C language fragments that were unsupported in previous work.

\end{abstract}

%% file: introduction.tex
\section{Introduction}

Many modern software systems comprise code written in a combination of programming languages \cite{10413900}. For example, the Robot Operation System 2 (ROS2) \cite{quigley2009ros,ros2} allows users to specify processes that communicate with each other, called ROS nodes, where each node can be implemented in a different programming language. Other languages and frameworks provide similar cross-language functionality for various domains, like Lingua Franca~\cite{lf} for real-time, distributed embedded systems, and Eclipse Cyclone DDS~\cite{eclipse-cyclonedds} and gRPC~\cite{indrasiri2020grpc} for distributed systems and networking. We refer to these multi-language software systems as {\em polyglot} systems.

Polyglot systems allow developers to pick and choose the best programming language for specific system components. Unfortunately, this flexibility comes at a cost, with studies indicating that the increased complexity of polyglot systems can lead to an increase in bugs \cite{10413900}. Recent work has addressed these issues through testing and other incomplete methods. For example, Yang et al. \cite{yang2024learning} develop a deep-learning-based method that helps localize bugs specifically due to the interaction between multiple languages in a single system. These tools and techniques can help developers build safer systems, but they cannot provide the guarantees needed in safety-critical settings and other vital settings in which polyglot systems appear. 

In this paper, we study the problem of formally verifying polyglot systems. Just like developing polyglot systems introduces new challenges in the development cycle, verifying polyglot systems introduces new challenges in the verification process. Over the past few decades, there has been substantial work on program verifiers that operate directly on 
individual programming languages. We refer to these tools as {\em language-specific verifiers}. For example, CBMC~\cite{cbmc2} and Frama-C~\cite{cuoq2012frama} can be used to verify C code, Kani~\cite{kani} and Verus~\cite{lattuada2023verus} for Rust,  ESC/Java~\cite{flanagan2002extended}, Java PathFinder~\cite{havelund2000model}, and JBMC for Java~\cite{cordeiro2018jbmc}, etc. However, very few approaches have been proposed for multi-language systems, so polyglot verification challenges remain largely unexplored.

The main existing approach for polyglot system verification translates all code to one common language and applies existing reasoning engines to that common representation. However, this approach is challenging due to the complexity of handling diverse syntax and semantics across languages, and arguably, it loses the benefits of having highly optimized language-specific verifiers. For instance, prior work, such as Lin et al. \cite{emsoft23LFVerifier}, only handles very restricted C code within the Lingua Franca language, highlighting the difficulties of such monolithic approaches. Encoding a multi-language system in a single verification model may also result in scalability issues, as the generated verification problem can become prohibitively large. To our knowledge, no existing approach enables automated verification of multi-language software systems through {\it cooperative} use of language-specific verifiers.

We propose a compositional verification approach 
to address this gap.
Our approach has two key components: (a) interface contracts written in an \textit{intermediate contract language} that abstract individual components, and (b) a {\it synthesizer} that automatically generates and refines these contracts with the help of language-specific verifiers.
%
The intermediate contract language is much smaller than the languages of individual components, allowing us to sidestep full language translation. 
Instead, we can use language-specific verifiers to check the contract validity by only translating the (much smaller) intermediate language fragment.
Further, our large language model (LLM)-based synthesizer and language-specific verifiers eliminate the burden of manually writing and refining contracts.
%
This enables scalable and automated verification of multi-language systems without requiring a full translation into a single verification framework.

In summary, the novel contributions of this paper are:
\begin{myitemize}    
\item \textbf{Polyglot verification problem.} We formalize the polyglot verification problem and propose a compositional solution integrating multiple language-specific verifiers. 
    
\item \textbf{Automated pre/post-condition contract synthesis.} We automate synthesis and refinement of pre/post-condition contracts using an intermediate language and combining counterexample-guided inductive synthesis (CEGIS) and abstraction refinement (CEGAR). Our approach uses synthesis oracles including syntax-guided synthesis (SyGuS) solvers and large language models (LLMs).

\item \textbf{Our tool.} We implemented \polyverl on top of UCLID5~\cite{polgreen-cav22,seshia-memocode18}, supporting C and Rust via CBMC~\cite{cbmc2} and Kani~\cite{kani}.
    Our implementation shows that an LLM-based synthesis oracle outperforms pure symbolic synthesis approaches.
    
\item \textbf{Case study and evaluation.} We developed a verifier for the Lingua Franca coordination language~\cite{lf} and show that our approach is able to verify multi-language systems with reasonable synthesis time.
    We also introduce a new set of polyglot verification benchmarks and realistic applications that are only solved by our approach.

\end{myitemize}

%% file: background.tex
\section{Motivation and Background} \label{sec:motivation}

%

\begin{figure}
    \centering
    \includegraphics[width=\linewidth]{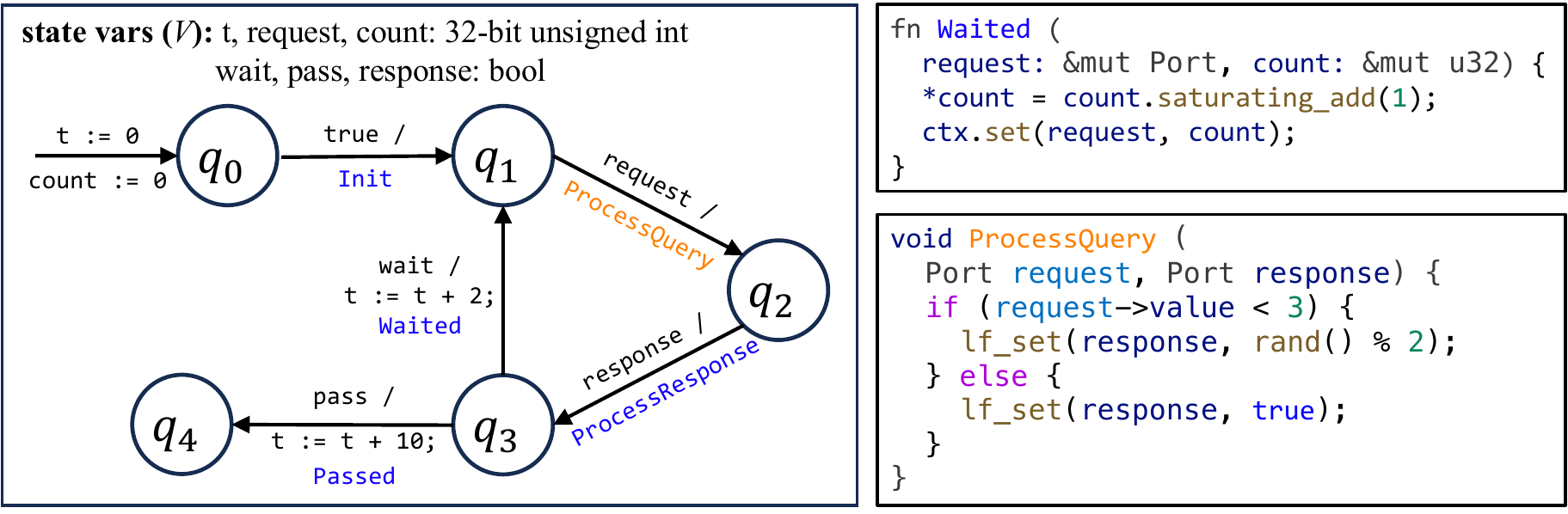}
    \caption{An Extended State Machine (ESM) describing a subway traffic control system implemented in Lingua Franca (LF). 
    $a/b$ means that when $a$ is present, $b$ is run.
    The procedures in \textcolor{blue}{\texttt{blue}} and \textcolor{orange}{\texttt{orange}} are Rust and C procedures respectively. Implementations of two procedures are shown on the right. \textsf{ctx.set} and \textsf{lf\_set} are LF API calls assigning values to ports.}
    \label{fig:fsm}
\end{figure}

%
Consider a train management system modeled in Figure~\ref{fig:fsm} as an Extended State Machine (ESM)~\cite{Alagar2011}, i.e., a guarded transition system where transitions have procedures attached to them.
Our example has procedures in two languages -- the \prwaited procedure in Rust and \prquery{} in C.

\begin{figure}
    \centering
    \includegraphics[width=\linewidth]{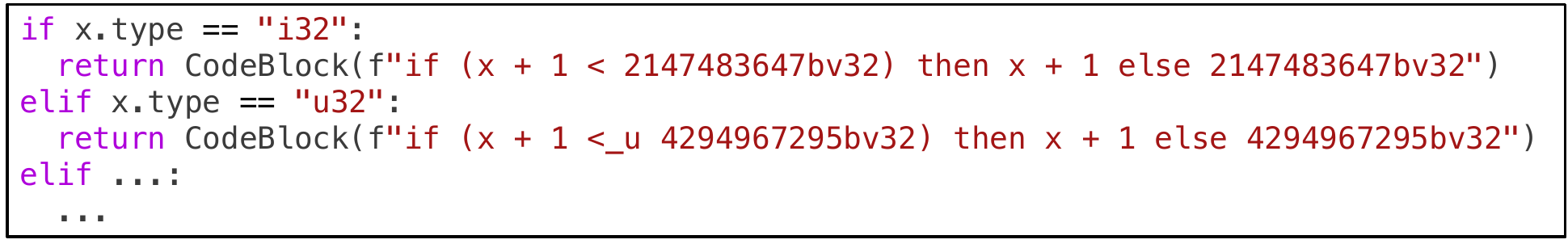}
    \caption{UCLID5 translation of \texttt{x.saturating\_add(1)} in Rust. Notice that the translation depends on the type of \texttt{x}.
    In UCLID5 and SMT-LIB, bitvectors do not have implicit type interpretation. Instead, they are only interpreted when specific operators are applied. For example, \texttt{<} (\texttt{bvslt}) and \texttt{<\_u} (\texttt{bvult}) stands for the signed and unsigned less than comparators in UCLID (SMT-LIB), respectively.}
    \label{fig:saturating_add}
\end{figure}

\smallskip
\noindent\textbf{Encodings must match language semantics.} 
%
Differences in language semantics can make encoding polyglot systems into a single formal model challenging. 
For example, the \texttt{x.saturating\_add(1)} Rust function in \prwaited saturates addition overflows to the maximum integer value (for that type) instead of wrapping around.
Figure \ref{fig:saturating_add} shows the corresponding UCLID5 expression of this function.
Encodings must faithfully capture \textit{all language-specific constructs} which can be large/complex for modern languages.

\smallskip
\noindent\textbf{Language-specific verification (LSV) tools} such as CBMC \cite{cbmc2} for C, Kani \cite{10.1145/3510457.3513031} for Rust, and Cameleer \cite{cameleer} for OCaml already perform this heavy lifting of faithfully capturing language semantics.
However, these tools might use different backends/proof techniques. For example, CBMC and Kani perform symbolic bounded model checking, while Cameleer uses deductive verification. 
%
Verifying polyglot systems with LSVs requires \textit{connecting these verifiers}, which is challenging due to these differences.

\smallskip
\noindent\textbf{Compositional (contract-based) verification} uses interface specifications, also called {\it contracts}, to abstract individual system components. Contracts can be assume-guarantee pairs for concurrent systems or pre/post-condition pairs for sequential software.
Once the validity of a contract w.r.t. the system component is checked using an LSV, it can be soundly used for system-level verification.
Crucially, given the contract, system-level check becomes \textit{agnostic of techniques/backends employed by the LSV}.
\polyverl uses this precise strategy of using contracts as the glue connecting language-specific verifiers.
%
In this sense, verification in \polyverl operates much like assume-guarantee/requires-ensures reasoning~\cite{jones1983tentative,liskov1986abstraction}. 
%
Manually identifying valid, yet \textit{strong enough} contracts, however, is challenging. 
E.g., \prquery{} (Fig. \ref{fig:fsm}) requires a non-trivial postcondition:

\begin{verbatim}
((request->value < 3) || response->value)
    && (request == \old(request))
\end{verbatim}

\smallskip
\noindent To address this, \textbf{\polyverl performs automated contract synthesis} by using refinement-based neural (LLM-based) contract search in combination with symbolic verification of the contract.
%
%
Thus, with contract synthesis, \polyverl bootstraps existing LSVs to solve the multi-language verification problem, at the same time avoiding the challenges of encoding the entire system into a single language.

%% file: problem_statement.tex
\newtheorem*{note}{Note}

\section{Problem Formulation}
\label{sec:problemformulation}

Polyglot systems are compositions of components where each component is a procedure in some language $\tl$ (e.g., C, Rust, Java).
Such a setup is widely seen in distributed systems where nodes execute code in a specific language and interact via mechanisms like remote procedure calls (RPCs).
We formalize such systems and define the verification problem.

We model polyglot systems as Extended State Machines (ESM)~\cite{leeseshia-book16}.  
%
Informally, ESMs capture inter-procedural control flow as transitions between \textit{modes}, where each transition has an associated \textit{update function} representing a sequence of procedure calls.
ESMs operate on a set of typed variables $V$ shared across all procedures in the system. 
We refer to the domain of (typed) variable $v \in V$ as $D(v)$ and denote the Cartesian product of the domains of all variables as $D(V)$. $D(V)$ thus forms the set of assignments $d$ to $V$. 
A {\it predicate} is a function 
$D(V) \rightarrow \mathbb{B}$. 
%
\begin{definition} [Extended State Machine] \label{def:efsm}
    An \emph{Extended State Machine (ESM)} is a tuple $(Q, V, I_0, T)$ where
    \begin{enumerate}
        \item $Q$ is a set of modes,
        \item $V$ is a set of typed variables,
        \item $I_0 \subseteq Q \times D(V)$ is a set of initial states,
        \item $T \subseteq (Q \times Q) \times (G \times U)$ is a set of mode-mode transitions, each associated with a guard in $G = D(V) \rightarrow \mathbb{B}$ and an update relation in $U = 2^{D(V) \times D(V)}$.
    \end{enumerate}
\end{definition}

\textbf{Executions.} The guard $g$ is a predicate determining whether a transition can occur, and the update relation $u$ defines how the transition affects $V$. At every step, the ESM in a state $(q, d)$ takes a transition $((q, q'), (g, u))$ such that $g(d)$ is true. The ESM then updates the state to $(q', d')$ where $(d, d') \in u$.
While the update relation ($u$) in each ESM transition is kept abstract, polyglot systems specialize them to be invocations of procedures (possibly from different languages).

\begin{definition} [Language]
    A \emph{language} $\tl$ is a set of terms and predicates constructed through syntactic rules over a set of variables $V$.
    The language fixes the interpretation of each term $t \in \tl$ and predicate $p \in \tl$ denoted as $\intp{t}_\tl: D(V) \rightarrow D(V)$ and $\intp{p}_\tl: D(V) \rightarrow \mathbb{B}$ respectively.
\end{definition}

We assume that the set $V$ contains all the variables read or written by the procedures implemented in the various languages.
%
Further, the interface variables shared across languages must have matching types, and hence we use the same $D(V)$ for different languages. This interface type-consistency is enforced in practical polyglot frameworks such as Lingua Franca \cite{lf} and ROS2 \cite{quigley2009ros}. 
Unless otherwise noted, we implicitly use the same variable set $V$ for the rest of the paper.
%

Next, we define procedures.

\begin{definition}[Procedure]             
    \label{def:external_procedure}
    A \emph{procedure} $f$ is a pair $(C, \tl)$ where $C \in \tl$ represents the procedure body. We call such a procedure an $\tl$-procedure.
    We refer to pre-state and post-state variables of a procedure as $V$ and $V'$, respectively.
    %
\end{definition}

We assume that procedures are total and always terminate
\textit{Polyglot models}, our model for polyglot systems, are ESMs where update relations are procedure invocation sequences.

\begin{definition} [Polyglot Model] \label{def:polyglot_model}
    A \emph{polyglot model} $\model$ is a tuple $(Q, V, I_0, T, F)$ where $(Q, V, I_0, T)$ is an ESM and $F$ is a set of procedures. Each transition $((q, q'), (g, u)) \in T$ is s.t. $u = \{(d, d') ~|~ d' = f_k \circ \cdots \circ f_1(d)\} $ for some $f_i \in F$, where $\circ$ denotes procedure composition: $f_1 \circ f_2(d) = \intp{C_1}_{\tl_1}(\intp{C_2}_{\tl_2}(d))$.
\end{definition}

An example of a polyglot model is shown in Figure~\ref{fig:fsm},
where $Q = \{q_i\}_i$, $I_0 = (q_0, \_)$ is the state after executing the procedures on the initial transition, $T$ is depicted by the transitions, and $F = \{\texttt{Init},~\texttt{ProcessQuery}, \cdots\}$.

\noindent 
{\underline{\textit{Remark on semantics:}}}
    While procedures are deterministic in our formulation (terms have deterministic semantics), nondeterministic procedures can be introduced by changing term interpretations from functions to relations.
    Further, side-effects are emulated by extending $V$ with global variables.
    %


\textbf{Traces.} Each execution of the model produces an infinite \textit{trace} that is the sequence of assignments ($d$) to variables in $V$ seen in the execution.
A property $\phi$ is a set of traces, typically in the form of a temporal logic formula.
Let the set of all traces that can be produced by a model $\model$ be $\mathcal{T}(\model)$.
We write $\model \models \phi$ if $\mathcal{T}(\model) \subseteq \phi$.
The polyglot verification problem aims to determine whether a polyglot model satisfies a property $\phi$.

\begin{problem_statement} [Polyglot verification problem]
    Given a polyglot model $\model = (Q, V, I_0, T, F)$ and a system property $\phi$ 
    determine whether $\model \models \phi$. 
    If $\model \not\models \phi$, provide a counterexample trace generated by $\model$ that violates $\phi$.
    \label{problem:polyglot}
\end{problem_statement}


\textbf{Verification Challenge and Insight.}
Crucially, while the procedure bodies ($C$) in $\model$ are known, \textit{capturing their interpretations ($\intp{C}_\tl$) is difficult}.
Naive monolithic verification of $\model$ would require encoding full semantics of $\tl$ (for all $\tl$) to a common modeling language, e.g. SMT-LIB.
This is extremely tedious and would require a parser/compiler infrastructure for each language. 
Our solution is to use contracts written in an \textit{intermediate language} $\il$ -- in this way we only need to support encodings for this (much smaller) language $\il$.

%% file: 4-verification-through-contracts.tex
\section{Polyglot Verification through a Contract IL}
\label{sec:polyglotcontracts}


We now introduce contract-based decomposition. 
%
%
%
The key component is a pre-post condition contract that abstracts (over-approximates) the behavior of a procedure $f$.

\begin{definition} [Contract] \label{def:contract}
    A ($\tl$-)\emph{contract} for procedure $(C, \tl)$ is a pair $(P, Q)$ where $P, Q$ are predicates in $\tl$.
    %
    $(P, Q)$ is \emph{valid} iff the Hoare-triple $\{P\}C\{Q\}$ holds, i.e.,
    \begin{equation} \label{eq:contract}
        \forall d, d' \in D(V). 
        (\intp{P}_\tl(d) \land d' = \intp{C}_\tl(d)) \implies \intp{Q}_\tl(d')
    \end{equation}
\end{definition}
%

By convention, preconditions $P$ can only refer to pre-state variables $V$, while postconditions $Q$ can refer to both pre- and post-state variables $V$ and $V'$.
Equation~\eqref{eq:contract} can be checked with a language-specific verifier for $\tl$.
%
A valid contract can serve as a sound system-facing abstraction of the procedure, as we now discuss.

\textbf{Contract composition} replaces procedures in a polyglot system $\model$ with their contracts.
\phantomsection
\label{sec:contract-composition}
%
%
Formally, for each transition $u = f_k \circ \cdots \circ f_1$ in $\model$, 
we generate contracts $(P_j, Q_j)$ for each $\tl_j$-procedure $f_j$ such that,
\begin{equation}
    \label{eq:contract-composition}
    \forall d\in D(V). \intp{Q_j}_{\tl_j}(d) \implies \intp{P_{j+1}}_{\tl_{j+1}}(d)
\end{equation}
Then, we define a new model $\model'$ where $u$ is replaced with the contract-derived relation $u' = \{(a, b) \mid \intp{P_1(a)}_{\tl_1} \implies \intp{Q_k(b)}_{\tl_{k}}\}$.
Since the replacement contract $u'$ overapproximates $u$, we have the following theorem.

\begin{theorem} \label{thm:soundness}
    If $\mathcal{M'} \models \phi$, then $\model \models \phi$.
\end{theorem}

\textbf{How does a contract IL help verification?} 
A key challenge with contract-based polyglot verification is checking Equation \eqref{eq:contract-composition}, which involves predicates \textit{in two different languages}.
We are unable to use a language-specific verifier to check this.
Our solution is to \textit{write contracts in an intermediate language} $\il$ and bridge contracts using a \textit{semantics-preserving compilation}. 

\begin{definition} [Intermediate Language]
    An \emph{intermediate language} (IL) $\il$ is a language of pre/post-condition contracts,
    whose semantics is defined by the SMT-based semantics of the language $\tl$ to which it is translated. 
\end{definition}

Concretely, we translate contract predicates from $\il$ to each procedure language $\tl$, e.g., C, Rust, through a semantics-preserving compilation $\compile_\tl$:

\begin{tcb}[$\il \rightarrow \tl$ Compilation]
    \label{tcb:compile}
\begin{equation*}
    \forall C\in \il.~ \intp{\compile_{\tl}(C)}_\tl \equiv \intp{C}_{\il}
\end{equation*}
\end{tcb}

Assumption~\ref{tcb:compile} allows us to compose and validate contracts in $\il$. 
When contracts are in $\il$, contract composition check (Equation~\eqref{eq:contract-composition}) simplifies to:
\begin{equation}
    \label{eq:contract-composition-il}
    \forall d\in D(V). \intp{Q_j}_{\il}(d) \implies \intp{P_{j+1}}_{\il}(d)
\end{equation}
Equation~\eqref{eq:contract-composition-il} can be checked by a model checker when $\il$ is translated into a modeling language like UCLID5.
Validity checking (Equation \eqref{eq:contract}) of an $\il$-contract $(P, Q)$ simply reduces to that of an $\tl$-contract $(\compile_{\tl}(P), \compile_{\tl}(Q))$.

\textbf{Counterexample guided contract refinement.} While Theorem \ref{thm:soundness} provides a soundness result, the converse does not hold.
A counterexample trace found in $\model'$ may be a spurious counterexample that cannot be produced by $\model$.
Contracts must be strengthened to exclude this trace.
However, developing \textit{valid contracts that are strong enough} to prove system-level properties is challenging, often involving iterative contract refinement.
To address this, we introduce the valid contract synthesis problem, which aims to generate a contract that avoids a set of spurious traces provided as \textit{negative} examples.
\begin{problem_statement}[Valid Contract Synthesis Problem] \label{problem:contract}
    Given an $\tl$-procedure $f$ and positive, negative example sets $\poseg, \negeg \subseteq D(V) \times D(V)$, synthesize a valid contract in $\il$ that includes the positive examples and excludes negative ones. 
\end{problem_statement}

\begin{figure}
    \centering
    \includegraphics[width=\linewidth]{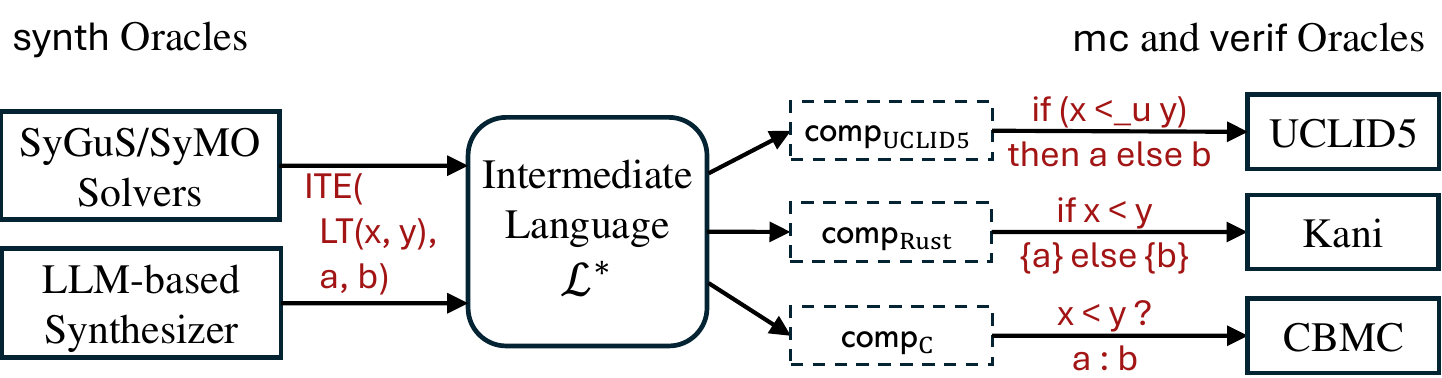}
    \caption{An intermediate language $\il$ act as glue between languages. Synthesis oracles generate expressions in $\il$, which are translated to semantic equivalent expressions in specific languages used by a model checker or verification oracles. $x$ and $y$ are unsigned bitvectors, hence using  \texttt{<\_u}  in UCLID5.}
    \label{fig:al}
\end{figure}

\textbf{How does a contract IL help synthesis?} 
Contract synthesis illustrates the second benefit of the intermediate language $\il$.
%
%
Language-specific contracts would require synthesizers tailored to the syntax and semantics of each language.
A common IL allows us to maintain a single interface with synthesis oracles. For instance, in the case of Syntax Guided Synthesis (SyGuS)~\cite{alur2013} we require a single grammar, or in the case of neural search (e.g., using LLMs), we require a single in-context example set.
Like verification, this greatly reduces the effort in developing synthesis oracles.
Figure~\ref{fig:al} illustrates how $\il$ mediates contract synthesis and verification.
%

%

%% file: approach.tex
\section{An Abstraction Synthesis Approach for Polyglot Model Verification}
\label{sec:approach}

We present an approach for contract synthesis (Problem \ref{problem:contract}) which we use as a subroutine to solve Problem \ref{problem:polyglot}.
We provide some background and then an overview in Section~\ref{subsec:overview}.

\subsection{Background: CEGIS and CEGAR}
\label{subsec:background}

We synthesize contracts by combining Counterexample Guided Inductive Synthesis (CEGIS)~\cite{DBLP:conf/asplos/Solar-LezamaTBSS06} and Counterexample Guided Refinement Abstraction (CEGAR)~\cite{DBLP:conf/cav/ClarkeGJLV00}.

CEGIS uses an iterative interaction between a \textit{learner} and a \textit{teacher} to synthesize a solution (in our case contracts) satisfying a set of constraints (in our case, example traces).
The \textit{learner} generates a candidate solution satisfying the constraints, while the \textit{teacher} checks whether the solution satisfies the overall specification. 
Synthesis succeeds if the solution is correct; else, the teacher provides a new counterexample.

CEGAR iteratively refines an abstraction (in our case contracts) in the context of a property. At each step, the abstraction is checked for correctness. If verification fails, the counterexample is analyzed to determine whether it is spurious (due to the abstraction being too coarse). If the counterexample is real, the system is buggy; else, the abstraction is refined.

\subsection{Approach Overview}
\label{subsec:overview}
\begin{figure}
    \centering
    \includegraphics[width=0.9\linewidth]{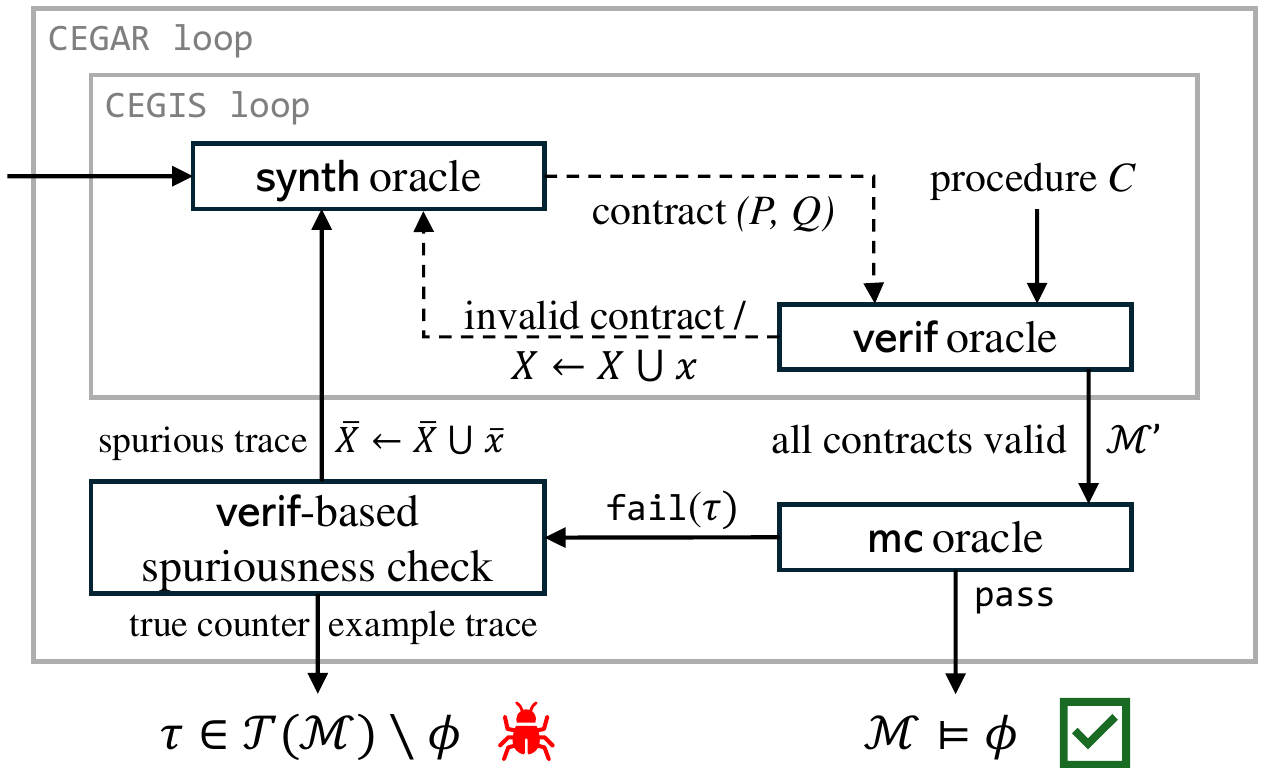}
    \caption{Flow diagram of our approach. The process starts from the synthesis oracle. Contracts are synthesized in the inner (dashed) CEGIS loop and refined in the outer CEGAR loop. $\poseg$ and $\negeg$ denotes positive and negative example sets.}
    \label{fig:diagram}
\end{figure}

We illustrate our CEGIS/CEGAR-hybrid approach to the polyglot verification in Figure~\ref{fig:diagram} and Algorithm~\ref{alg:polyver}. 
The approach is parameterized by the set of synthesis oracles, verification oracles, and the intermediate contract language $\il$.
It operates by performing CEGIS (inner loop; line~\ref{lst:line:cegis}) using synthesis and verification oracles to synthesize contracts for each procedure, which solves Problem~\ref{problem:contract}.
CEGAR (outer loop) then uses these contracts to generate the induced \hyperref[sec:contract-composition]{abstract model} $\model'$ (line~\ref{lst:line:induce}) specified entirely in $\il$ by combining the system model and the contracts. 
We then check these contracts against the system-level property (line~\ref{lst:line:mc}) -- if the property is proven, the system is verified (line~\ref{lst:line:verified}); otherwise, we check if the counterexample trace is spurious (line~\ref{lst:line:check_spurious}). 
If the trace is spurious, we exclude it by returning to the inner CEGIS loop to refine contracts, else we report failure with the trace (line~\ref{lst:line:fail}).
Thus, \polyverl iteratively switches between CEGIS and CEGAR to generate a counterexample or prove the property.

\begin{algorithm}
\caption{PolyVer: CEGIS/CEGAR-hybrid approach}
\label{alg:polyver}
\begin{algorithmic}[1]
\Require $\model = (Q, V, I_0, T, F)$: a polyglot model, $\phi$: a property
\Ensure \textit{res}: verification result, $\tau$: a counterexample trace
\ForAll{$f_i \in F$}
    \State $\poseg_i, \negeg_i \gets \emptyset, \emptyset$
\EndFor
\While{$\top$} \Comment{CEGAR loop}
    \State $\mathcal{C} \gets \{ \text{synthContract}(f_i, \poseg_i, \negeg_i)\}_{f_i}$ \Comment{CEGIS loop}\label{lst:line:cegis}
    \State $\mathcal{M'} \gets \text{induce}(\model, \mathcal{C})$ \label{lst:line:induce}
    \State $\textit{res} \gets \modelcheck(\mathcal{M'}, \phi)$ \label{lst:line:mc}
    \If{$\textit{res} = \texttt{pass}$} \Comment{Verification succeeds}\label{lst:line:verified}
        \State \Return \texttt{pass}, -
    \ElsIf{$\textit{res} = \texttt{fail}(\tau)$}
        \State $\textit{isSpurious} \gets \bot$
        \ForAll{$f_i \in F$}
            \State $\negeg_i' \gets \text{checkSpurious}(\tau, f_i)$\label{lst:line:check_spurious}
            \If{$\negeg_i' \neq \emptyset$}
                \State $\textit{isSpurious} \gets \top$
            \EndIf
            \State $\negeg_i \gets \negeg_i \cup \negeg_i'$
        \EndFor
        \If{$\neg \textit{isSpurious}$}\Comment{Counterexample found}
            \State \Return \texttt{fail}, $\tau$ \label{lst:line:fail}
        \EndIf
    \EndIf
\EndWhile
\end{algorithmic}
\end{algorithm}


\subsection{Oracles}
We assume access to synthesis and verification oracles, and a model checker.
Synthesis oracles accept positive, negative example sets $\poseg$, $\negeg$,
and return an $\il$-contract $(P, Q)$ that satisfies $\poseg$ and dissatisfies $\negeg$:
\begin{equation*}
    \synthoracle(\poseg, \negeg, C \in \tl) \rightarrow (P \in \il, Q \in \il)
\end{equation*}

A verification oracle call $\verifyoracle(P, Q, C)$ verifies $\il$-contract validity w.r.t. an $\tl$-procedure (producing a \textit{positive}-example if invalid):
The verifier is a part of our trusted computing base (TCB) and is assumed to be sound and complete:
\begin{tcb}[Verifier]
    \label{tcb:verifier}
    \begin{align*}
        \verifyoracle(P, Q, C) &= \texttt{pass} \iff \{P\} ~ C ~ \{Q\} \\
        \verifyoracle(P, Q, C) &= \texttt{fail}(\v{x} = (d, d')) \iff \\
        &\intp{P}_\il(d) \land d' = \intp{C}_\tl(d) \land \neg \intp{Q}_\il(d')
    \end{align*}
\end{tcb}

A model checker call $\modelcheck(\model'[P_i, Q_i], \phi)$ verifies the abstract model $\model'$ (with contracts) w.r.t. property $\phi$, and produces a (\textit{negative}-) counterexample trace $\tau$ if violated.
Examples $Y_f$ can be extracted from $\tau$ for each procedure $f$:
\begin{tcb}[Model Checker]
    \label{tcb:modelchecker}
    \begin{align*}
        &\modelcheck(\model'[P_i, Q_i], \phi) = \texttt{pass} \iff \model' \models \phi \\
        &\modelcheck(\model'[P_i, Q_i], \phi) = \texttt{fail}(\v{\tau}) \iff \v{\tau} \in \mathcal{T}(\model') \setminus \phi \\
        & \qquad \text{with } \mathsf{extract}(\tau) = \{Y_f\}_f
    \end{align*}
\end{tcb}

We now detail contract synthesis and refinement in Sections~\ref{subsec:cegis} and \ref{subsec:spurious_check} respectively.
Finally, we discuss correctness guarantees in Section~\ref{subsec:correctness}.

\subsection{Contract Synthesis via CEGIS Loop} 
\label{subsec:cegis}
The (inner) CEGIS loop iterates to generate a valid contract for procedure $(C, \tl) \in F$, described in Algorithm~\ref{alg:synthesize_contract}.
%
In each iteration, given $C \in \tl$, and initially empty example sets $\poseg$ and $\negeg$, 
$\synthoracle(\poseg, \negeg, C)$ generates a contract $(P, Q)$ in $\il$.
Then, $\verifyoracle$ checks contract validity with respect to the procedure $C$.
If $\verifyoracle(P, Q, C) = \texttt{pass}$ the contract is valid;
otherwise the positive example $\v{x}$ is added to $\poseg$ and the loop continues until a valid contract is found.
Under Assumption~\ref{tcb:verifier}, this produces valid contracts upon termination, thereby solving Problem~\ref{problem:contract}.

\begin{algorithm}
\caption{$\text{synthContract}(f, \poseg, \negeg)$}
\label{alg:synthesize_contract}
\begin{algorithmic}[1]
\Require $f = (C, \tl)$: a procedure
\Modify $\poseg, \negeg$: a set of positive/negative examples
\Ensure $(P, Q)$: a valid contract for $f$
\While{$\top$}
    \State $(P, Q) \gets \synthoracle(\poseg, \negeg, C)$
    \State $\textit{res} \gets \verifyoracle(P, Q, C)$
    \If{$\textit{res} = \texttt{pass}$} \Comment{Contract is valid}
        \State \Return $(P, Q)$
    \ElsIf{$\textit{res} = \texttt{fail}(x)$} \Comment{Contract is invalid}
        \State $\poseg \gets \poseg \cup \{x\}$ \Comment{Collect positive example}
    \EndIf
\EndWhile
\end{algorithmic}
\end{algorithm}


\newtheorem{example}{Example}

\begin{example}[Contract synthesis] 
\label{ex:cegis}
\emph{
    The below table illustrates CEGIS-based contract synthesis for \prquery{}.
    $\synthoracle$ synthesizes $(P_1, Q_1)$ which $\verifyoracle$ checks for validity returns a positive example $(d, d')$. Note that the pre-state $d$ satisfies $P_1$ but the post-state $d'$ violates $Q_2$. $\synthoracle$ then refines the contract based on $(d, d')$ and generates $(P_2, Q_2)$, which is again checked by $\verifyoracle$. \texttt{req} and \texttt{resp} are short for \texttt{request} and \texttt{response} respectively.
}
    \begin{table}[ht]
        \centering
        \begin{tabular}{l|c}
            Oracle & Output \\ \hline
            $\synthoracle$ iter. 1 & $P_1: \top, Q_1:(\texttt{resp == (req >= 3))}$ \\ \hline
            $\verifyoracle$ iter. 1 & $\texttt{fail}\ /\ (\texttt{req}, \texttt{resp}):d = (\texttt{1}, \bot), d' = (\texttt{1}, \top)$ \\ \hline
            $\synthoracle$ iter. 2 & $P_2:\top, Q_2:\texttt{(req < 3 || resp == $\top$)}$ \\ \hline
            $\verifyoracle$ iter. 2 & $\texttt{pass}\ /\ -$ \\
    \end{tabular}
\end{table}
\end{example}

Valid contracts are then used to construct $\model'$ (\S\ref{sec:contract-composition}) which is then checked by $\modelcheck$.
This may either succeed or return a counterexample trace, which is then checked for spuriousness.

\subsection{Spuriousness Check} \label{subsec:spurious_check}
A counterexample trace generated by $\modelcheck$ may be \textit{spurious}, i.e., unreproducible by the procedures.
We illustrate the checking process in Algorithm~\ref{alg:check_spurious}.
We first extract pre/post variable assignments $\v{p} = \{(d, d')\}_i$ for each procedure call in the trace (line~\ref{lst:line:extract}).
For each $(d, d') \in \v{p}$, we check whether $\{V = d\}~ C ~\{V' \neq d'\}$ holds using $\verifyoracle$. 
If it holds, the pair cannot be produced by the procedure and is then added as a negative example to $\negeg$ (``spurious trace'' edge in Figure~\ref{fig:diagram}; line~\ref{lst:line:neg}). 
This will be excluded during the next CEGIS loop refinement iteration.
If all pre-post transitions are valid for all procedures, the trace is a true counterexample (``true cex trace'' edge). 
%
%
Example~\ref{ex:cegar} shows an example of a spuriousness check performed on \prquery{} against a pair of assignments.

\begin{algorithm}
\caption{$\text{checkSpurious}(\tau, f)$}
\label{alg:check_spurious}
\begin{algorithmic}[1]
\Require $\tau$: a counterexample trace, $f = (C, \mathcal{L})$: a procedure
\Ensure $\negeg$: a set of negative examples for $f$
\State $\negeg \gets \emptyset$
\ForAll{$(d, d') \in \mathsf{extract}(\tau)$} \label{lst:line:extract}
    \State $\textit{res} \gets \verifyoracle(V = d, V' \neq d', C)$
    \If{$\textit{res} = \texttt{pass}$} \Comment{$\tau$ is spurious}
        \State $\negeg \gets \negeg \cup \{(d, d')\}$ \Comment{Collect negative example}\label{lst:line:neg}
    \EndIf
\EndFor
\State \Return $\negeg$
\end{algorithmic}
\end{algorithm}

\begin{example}[Spuriousness check]
\label{ex:cegar}
\emph{
    Continuing from Example~\ref{ex:cegis}, assume that a model checker ($\modelcheck$) checks the property using $(P_2, Q_2)$ and returns a trace containing $d$ and $d'$. 
    To check spuriousness, we check the Hoare-triple $\{(\texttt{req}, \texttt{resp}) = d\}$ \prquery{} $\{(\texttt{req}', \texttt{resp}')\neq d'\}$ using $\verifyoracle$, who returns \texttt{pass}. 
    The pair is thus impossible, making the trace spurious. We then return to the CEGIS loop and re-synthesize a contract, which is proven valid by $\verifyoracle$.
}
    
    \begin{table}[ht]
        \centering
        \begin{tabular}{l|c}
        Oracle & Output \\ \hline
        $\modelcheck$ in CEGAR & $(\cdots, d = (\texttt{2}, \top), d' =(\texttt{10}, \top), \cdots)$ \\ \hline
        $\verifyoracle$ in CEGAR & $\texttt{pass}\ /\ (\texttt{req}, \texttt{resp}) : d = (\texttt{2}, \top), d' = (\texttt{2}, \top)$ \\ \hline
        $\synthoracle$ in CEGIS & $P_3:\top, Q_3:\texttt{(req < 3 || resp == $\top$)}$ \\
        & \texttt{\&\& \textbackslash old(req) == req} \\ \hline
        $\verifyoracle$ in CEGIS & $\texttt{pass}\ /\ -$ \\
    \end{tabular}
\end{table}
\end{example}

\subsection{Correctness Guarantee} 
\label{subsec:correctness}
Our approach makes assumptions \ref{tcb:compile}, \ref{tcb:verifier}, and \ref{tcb:modelchecker}.
%
Given these assumptions, we guarantee \textit{partial correctness}, meaning that if \polyverl terminates with \texttt{pass} result, the system is verified, $\model \models \phi$ (soundness), and if it terminates with a \texttt{fail} result, the system is unsafe, $\model \not\models \phi$ (completeness).

We intuit this result as follows.
For soundness, notice that contracts used in the induced model must overapproximate procedure behaviors because of the verification oracle assumption (Assumption \ref{tcb:verifier}).
This means that when $\modelcheck$ reports \texttt{pass}, there does not exist a trace that is producible by the procedures that violates the property, since all producible behaviors have already been captured by the contracts.
For completeness, when $\modelcheck$ reports \texttt{fail} with a trace, the trace must violate the property.
This trace must also be producible by the procedures by the spuriousness check (\S\ref{subsec:spurious_check}).

%% file: implementation.tex
\section{\polyverl Implementation}
We implement the \polyverl automated approach in a tool\footnote{Available at \hyperlink{https://github.com/uclid-org/uclid}{https://github.com/uclid-org/polyver}.} which builds on top of the UCLID5 verifier~\cite{polgreen-cav22} and currently supports 
the C and Rust languages.
%
%
The tool is modular and can be extended to support other languages and verifiers.
Users need to provide oracles that subscribe to certain interfaces, which we now describe.
We also illustrate oracles we have used for C and Rust.
We use Boolean connectives, arithmetic operations, 
and comparators for the intermediate language $\il$.


\subsection{Synthesis Oracle Interface}
Given a procedure $C$, a set of positive and negative examples $\poseg, \negeg$, 
the synthesis oracle $\synthoracle$ returns an $\il$-contract that is consistent with the examples.
Below we discuss two candidate synthesis oracles and their interfaces.

\subsubsection{LLM-based Synthesizer}
\label{subsubsec:llmsynth}
LLMs are a good fit for contract synthesis because of their understanding and reasoning abilities across multiple (common) programming languages.
We developed an LLM-based synthesizer that uses chain-of-thought prompting \cite{wei2022chain} with error and example feedback to generate contracts, i.e., the LLM is asked to first summarize the behavior of $C$ and then generate a contract in $\il$.

Inspired by work by Mora et al.~\cite{moraNeurIPS2024}, we formulate a Python-based DSL that allows the LLM to construct $\il$-expressions in a familiar language.
We note that this does not affect the expressivity of generated the contract expressions.
However, it increases the LLM's ability to generate syntactically correct expressions.
If the expression is invalid, we re-prompt the LLM with an error message indicating incorrect syntax or inconsistent types. 
Provided with positive and negative examples, the LLM is asked to find a constraint that distinguishes the two sets. 
%
Details of our prompt are provided in Appendix~\ref{appendix:llm}. 


\subsubsection{SyGuS/SyMO-based Synthesizer}
The second oracle treats the problem as a Programming by Example (PBE) problem \cite{gulwani2016programming}, asserting that the generated contract $(P, Q)$ must\footnote{Note that unlike Assm. \ref{tcb:verifier}, this is not a part of the trusted base of the \polyverl approach. In particular, the LLM-based synthesizer (Sec. \ref{subsubsec:llmsynth}) may not satisfy this.} satisfy examples in $\poseg$ and dissatisfy those in $\negeg$:
\begin{align*}
    &\forall (d, d') \in \poseg.~~ \intp{P}_\il(d) \implies \intp{Q}_\il(d') \\
    &~~\land~~ \forall (d, d') \in \negeg.~~ \intp{P}_\il(d) \land \neg \intp{Q}_\il(d')
\end{align*}
%
%

We use Syntax-Guided Synthesis (SyGuS)~\cite{alur2013} solvers to solve the PBE problems.
A similar approach is to encode as a SyGuS Modulo Oracles (SyMO)~\cite{DBLP:conf/vmcai/PolgreenRS22} problem, a variant of SyGuS that allows constraints to contain black-box functions.
%
We use SyMO solvers with language-specific verifiers as black-box functions.
As the solver itself forms a CEGIS loop, we can directly solve Problem~\ref{problem:contract} in a single iteration.


\subsection{Verification Oracle Interface}
We use CBMC and Kani as verification oracles for C and Rust, respectively.
Given a procedure body $C$ and a contract $(P, Q)$, we generate CBMC and Kani queries by constructing $\{P\}C\{Q\}$ using their respective APIs.
Listing~\ref{lst:ums-cbmc} shows an example of a CBMC query.
Specifically, at lines 20 and 22, we use CBMC APIs to assume and assert the \texttt{PRECONDITION} and \texttt{POSTCONDITION} C macros around the function call at line 21.
This query is then verified using CBMC.
We apply a similar template for Rust procedures, verified with the Kani model checker.
CBMC and Kani can achieve unbounded correctness for programs with loops by using loop invariants or loop unwinding assertions.


\begin{lstlisting}[language=C, caption=CBMC query generated for \texttt{ProcessQuery}, label=lst:ums-cbmc]
// ... Definitions of structs and APIs ...
void ProcessQuery() {
    if (query->value < 3) { 
        lf_set(response, rand() % 2 == 1);
    } else { lf_set(response, true); }
}
int main() {
    // calloc for inputs and the self struct.
    query = calloc(1, sizeof(UMS_query_t));
    response = calloc(1, sizeof(UMS_response_t));
    // Assume that there are no NULL pointers.
    __CPROVER_assume(query && response);
    // Initialize with nondeterministic values.
    *query = nondet_UMS_query_t();
    // CBMC checks pre/post-conditions.
    __CPROVER_assume(PRECONDITION);
    ProcessQuery();
    assert(POSTCONDITION);
}

\end{lstlisting}

%% file: case_study.tex
\section{Case Study: LF Program Verification}
We demonstrate \polyverl's effectiveness by developing a verifier for the Lingua Franca (LF) coordination language \cite{lf}.
%
LF is a polyglot coordination language designed for real-time, concurrent, and distributed systems.
LF programs contain concurrent components called \textit{reactors} connected through \textit{ports} and \textit{connections}. 
A reactor contains reactive message handlers called \textit{reactions}, which can be written in a target programming language, e.g., C, C++, Rust, Python, or TypeScript. 
A reaction is invoked when any of its triggers are present, and it may produce effects triggering downstream reactions.
With multiple reaction languages, Lingua Franca verification exhibits all the polyglot system verification challenges discussed previously.
%
\begin{figure*}[t]
\centering
\begin{minipage}{\textwidth} 
\begin{multicols}{2}
\begin{lstlisting}[basicstyle=\ttfamily\scriptsize,style=lfStyle]
// Rust reactor
reactor Train {
  input response : bool
  output request : u32
  state count : u32 = 0
  state passed : bool = false
  logical action wait(2 minute)
  logical action pass(10 minute)

  reaction Init (startup) -> request {= 
    ctx.set(request, self.count);
  =}
  reaction Waited (wait) -> request {=
    self.count = self.count.saturating_add(1);
    ctx.set(request, self.count);
  =}
  reaction Passed (pass) {= self.passed = true; =}
  reaction ProcessResponse (response) -> pass, wait {=
    if ctx.get(response).unwrap() {
      ctx.schedule(pass, 0);
    } else { 
      ctx.schedule(wait, 0);
    }
  =}
}
\end{lstlisting}
\begin{lstlisting}[basicstyle=\ttfamily\scriptsize,style=lfStyle,firstnumber=26]
// C reactor
reactor UMS {
  input request : uint32_t
  output response : bool

  reaction ProcessQuery (request) -> response {=
    if (request->value < 3) {
      lf_set(response, rand() % 2);
    } else { lf_set(response, true); }
  =}
}

// Property to be verified
@property(
  name="pass_within_12_seconds",
  spec="F[0, 12 minute](t.passed)",
)
main reactor {
  // Component instantiations
  u = new UMS()
  t = new Train()
  // Component interactions
  t.request -> u.request
  u.response -> t.response
}
\end{lstlisting}
\end{multicols}
\end{minipage}
\caption{LF program for a subway control system from~\cite{halbwachs1992programming}.}
\label{fig:lf-train}
\end{figure*}


\subsection{Our Running Example as a Lingua Franca Program} \label{subsec:running}
Figure~\ref{fig:lf-train} shows the full LF program for our running example (Fig. \ref{fig:fsm}), a subway U-turn management system (UMS).
%
%
The \texttt{Train} and \texttt{UMS} reactors, written in Rust and C respectively, use special brackets \texttt{\{=...=\}} to enclose reaction bodies.
The system describes a train asking for permission to cross.
The train sends out requests to the UMS, which decides whether to grant permission (\texttt{ProcessQuery}).
If the train has previously made more than three attempts, the UMS will certainly grant permission; otherwise permission is granted randomly.
If granted, the train will pass, taking ten minutes; otherwise, it waits two minutes before retrying (\texttt{ProcessResponse}).
The property ($\phi$) that the train passes in 12 minutes is false, since the train may take up to 16 minutes.



\subsection{Methodology: Mapping LF Programs to Polyglot Models} \label{subsec:mapping}
LFVerifier~\cite{emsoft23LFVerifier} performs end-to-end LF verification by mapping the program onto a State Space Diagram (SSD) and compiling it along with the reaction body (C code) into a monolithic axiomatic model.
\polyverl instead compiles the SSD into an operational polyglot model.


An SSD is a graph where each node (state) $s$ contains the current logical time, a set $R$ of reactions to be executed, and a set of pending events,
and each edge represents logical time advancement. 
Figure~\ref{fig:ssd} in the Appendix shows an example SSD generated from the LF program in Figure~\ref{fig:lf-train}. 

LF timing semantics at each state $s$ impose a directed-acyclic graph (DAG) over the set of reactions $R$, with edges encoding execution precedence. 
A reaction may fire only when all of its predecessors have run to completion, so any topological sort of the DAG is a valid execution order. 
Independent reactions whose predecessors have finished may execute in either order, hence introducing nondeterministic reaction interleaving. 
Once all reactions in $R$ are completed, the SSD state can advance to the next state.


We encode time progression and nondeterministic reaction interleaving in UCLID5 by encoding reaction DAGs and dynamically tracking their execution status.
%
This yields a nondeterministic polyglot model (Definition~\ref{def:polyglot_model}) where each transition executes one reaction, as in the model in Figure~\ref{fig:fsm}.
We integrate the \polyverl-based verifier into the LF compiler, which generates C, Rust, and UCLID5 code and performs verification.

%% file: evaluation.tex
\section{Experimental Evaluation} \label{sec:evaluation}

Our experimentation evaluates \polyverl on both its components: contract synthesis and verification using synthesized contracts. We want to answer the following research questions:
%

\noindent \textbf{RQ1.} How does \polyverl compare to prior LF verification work? \textbf{RQ2.} Can \polyverl handle full-fledged LF examples? \textbf{RQ3.} Can \polyverl verify multi-language systems? 

\textbf{Experiment Setup.}
We run all experiments on a single core on a 3.7GHz Intel Core i9 machine with 62GB RAM.
We generate contracts using an LLM-based synthesizer (OpenAI o1-mini) with three parallel LLM queries per procedure, and use CBMC-6.3.1, Kani-0.56.0, and z3-4.8.7 for verification.
Parallel synthesis for multiple procedures is implemented in \polyverl but not used in the experiments.
Since LF reactions often include unmodified variables irrelevant to strong/tight contracts, our LLM prompt includes a single instruction emphasizing the importance of identifying these variables.

\begin{table}
    \centering
    \caption{The synthesis oracle time (SOT) is the total contract synthesis time for all procedures (proc) across all CEGIS iterations (IS). The verification oracle time (VOT) is the total time on verification oracles, including contract validation and spuriousness checks. The verification time (UT) is the total time on UCLID5, called once per CEGAR iteration (AR).}
    \setlength{\tabcolsep}{0.1cm}
    \begin{tabular}{l|c|cc|cc|ccc|c}
        \toprule
        & & \multicolumn{2}{c}{\textbf{LOC}} & \multicolumn{2}{c}{\textbf{\#Iters}} & \multicolumn{4}{c}{\textbf{CPU Time (s)}}  \\
        \textbf{Benchmark} & \#proc & C & LF & IS & AR & SOT & VOT & UT & \cite{emsoft23LFVerifier} \\
        \hline
        \texttt{ADASModel} & 7 & 16 & 73 & 7 & 1 & 220.9 & 3.2 & 21.5 & 24.1 \\ 
        \texttt{AircraftDoor} & 3 & 10 & 38 & 3 & 1 & 82.1 & 1.2 & 2.0 & 6.1 \\
        \texttt{Alarm} & 2 & 12 & 30 & 2 & 1 & 62.3 & 0.9 & 2.1 & 5.8 \\
        \texttt{CoopSchedule} & 2 & 2 & 37 & 2 & 1 & 54.5 & 0.8 & 13.4 & 15.5 \\
        \texttt{Election} & 3 & 7 & 35 & 3 & 1 & 124.2 & 1.4 & 83.9 & 35.7 \\ 
        \texttt{Election2} & 2 & 6 & 31 & 3 & 2 & 142.8 & 1.5 & 11.7 & 11.4 \\
        \texttt{Elevator} & 13 & 53 & 206 & 13 & 1 & 413.1 & 33.4 & 664.8 & 106.1 \\
        \texttt{Factorial} & 3 & 11 & 36 & 3 & 1 & 138.2 & 1.9 & 63.0 & 42.4 \\ 
        \texttt{Fibonacci} & 5 & 14 & 39 & 5 & 1 & 167.3 & 2.4 & 390.9 & 264.3 \\ 
        \texttt{PingPong} & 4 & 8 & 62 & 4 & 1 & 131.0 & 1.8 & 8.2 & 12.0 \\
        \texttt{Pipe} & 5 & 10 & 60 & 5 & 1 & 147.4 & 2.4 & 172.2 & 144.1 \\ 
        \texttt{ProcessMsg} & 4 & 10 & 32 & 4 & 1 & 116.6 & 1.8 & 5.1 & 14.6 \\
        \texttt{ProcessSync} & 1 & 2 & 18 & 1 & 1 & 35.2 & 0.4 & 1.7 & 5.0 \\
        \texttt{Railroad} & 10 & 64 & 66 & 17 & 2 & 678.9 & 9.0 & 264.8 & 223.4 \\
        \texttt{Ring} & 4 & 7 & 45 & 4 & 1 & 105.6 & 1.8 & 6.8 & 16.7 \\
        \texttt{RoadsideUnit} & 8 & 26 & 81 & 8 & 1 & 259.0 & 3.8 & 25.3 & 44.0 \\
        \texttt{SafeSend} & 4 & 12 & 53 & 4 & 1 & 124.3 & 1.8 & 2.3 & 6.9 \\
        \texttt{Subway} & 10 & 20 & 64 & 10 & 1 & 266.5 & 4.5 & 67.2 & 35.0 \\ 
        \texttt{Thermostat} & 6 & 29 & 45 & 6 & 1 & 192.6 & 2.6 & 3.5 & 12.8 \\
        \texttt{TrafficLight} & 4 & 46 & 42 & 8 & 2 & 462.1 & 4.4 & 216.1 & - \\
        \texttt{TrainDoor} & 3 & 7 & 50 & 3 & 1 & 74.5 & 1.2 & 2.1 & 5.8 \\
        \texttt{UnsafeSend} & 4 & 12 & 53 & 4 & 1 & 122.1 & 1.8 & 2.9 & 6.8 \\
        \bottomrule
    \end{tabular}
    \label{tab:benchmark_data}
\end{table}

\subsection{RQ1: Comparison to prior LF verification work}
\label{subsec:rq1}

We contrast \polyverl with LFVerifier~\cite{emsoft23LFVerifier} using its proposed LF benchmarks.\footnote{While these benchmarks may be present in the LLM training data, no contracts are available.}
%
These are arguably toy examples due to the limited C syntax~\cite{emsoft23LFVerifier} supports  -- only assignments, conditionals (incomplete), and arithmetic and Boolean operations. 
They use a subset of safety MTL properties~\cite{10.1007/11691372_27}.
A direct comparison is not entirely fair since \cite{emsoft23LFVerifier} uses a monolithic, axiomatic encoding specialized for LF programs, while \polyverl uses a compositional, operational encoding that is more general.
%
%
Despite this, \cite{emsoft23LFVerifier} is, to our knowledge, the only prior work that performs end-to-end automated LF program verification, so we compare with it.
%


We report on benchmark details, synthesis iteration counts, and runtimes in Table~\ref{tab:benchmark_data}.
\polyverl successfully synthesizes the contracts and verifies \textit{all} benchmarks automatically, while \cite{emsoft23LFVerifier} is unable to verify the \texttt{TrafficLight} example.
In 18 out of the 22 benchmarks, the contracts were synthesized correctly in a single CEGIS loop, taking 37 seconds on average.
For the rest, we observe that all but one reaction still requires one CEGIS iteration while one (hard) reaction requires all the remainder of iterations. 
The hard reaction often includes complex control flow and data dependencies, which makes it difficult even for human experts to reason about.

We record the synthesis oracle time (SOT), verification oracle time (VOT), and verification runtime (UT).
%
Though the total runtime (SOT + VOT + UT), dominated by the contract synthesis time, is generally slower than that of \cite{emsoft23LFVerifier}, we argue that this is time well spent as \polyverl offers a significant \textbf{qualitative improvement} over \cite{emsoft23LFVerifier}:
There is no longer a need to develop a translator from a target language to the modeling language (e.g., C to UCLID5). 
Since we leverage robust language-specific verifiers (e.g., CBMC, Kani), we are also able to support a much more comprehensive fragment of C/Rust. 
For example, CBMC officially supports ANSI-C, while \cite{emsoft23LFVerifier} develops a custom C-to-UCLID5 translator for a small C subset (which is what these benchmarks use).
These benefits are highlighted in Section~\ref{subsec:rq2} and \ref{subsec:rq3}.
%

We also experimented with SyGuS/SyMO solvers but found that these approaches required a large number of CEGAR iterations and failed to terminate in most benchmarks. 
We attribute this to the solvers' lack of access to procedure bodies, relying solely on positive and negative examples for guidance.

\subsection{RQ2: Can \polyverl handle full-fledged LF examples?}
\label{subsec:rq2}

%
We now evaluate \polyverl on examples requiring a broader fragment of C  that \cite{emsoft23LFVerifier} is unable to verify.
We evaluate our approach on two embedded systems -- a blinking LED controller and a hill-climbing robot -- and a satellite altitude controller~\cite{cardoso2022reactionwheel,lin2024pretvm}.
The LED controller tracks blinking status, while the robot checks if it is in climbing mode when on a ramp.
Both are resource-constrained embedded systems.
The satellite example uses a PID controller and disables motors if user input exceeds safe limits.
When library functions (e.g., \texttt{atan}) are unavailable, we overapproximate them using nondeterministic input/output values.


The results in Table~\ref{tab:hard} show that our approach is able to handle language complexities and generate contracts for actual implementation code.
In the LED and hill climbing examples, the LLM-based synthesizer was able to summarize procedure behavior while ignoring details about low-level code when they are irrelevant.
In the satellite example, the LLM struggled to synthesize a valid contract for the PID controller.
However, we noticed that providing a simple comment in the code that captures the logic of the controller greatly helped synthesis.
The synthesized contracts can all be checked by CBMC.
This highlights \polyverl's potential to verify real-world systems.

\begin{table}
    \centering
    \caption{Results for full-fledged and multi-language examples. The Blink, HillClimb, and Satellite examples include real implementation C code. The rest and other 22 variants in Table~\ref{tab:benchmark_data} are multi-language systems only solvable by \polyverl.}
    \setlength{\tabcolsep}{0.1cm}
    \begin{tabular}{l|c|cc|cc|ccc}
        \toprule
        & & \multicolumn{2}{c}{\textbf{LOC}} & \multicolumn{2}{c}{\textbf{\#Iters}} & \multicolumn{3}{c}{\textbf{CPU Time (s)}}  \\
        \textbf{Benchmark} & \#proc & C/Rust & LF & CEGIS & CEGAR & SOT & VOT & UT \\
        \hline
        \texttt{Blink} & 1 & 123 & 35 & 1 & 1 & 31.0 & 0.45 & 1.26 \\
        \texttt{HillClimb} & 12 & 75 & 102  & 20 & 1 & 685.4 & 74.9 & 9.7 \\
        \texttt{Satellite} & 6 & 424 & 166 & 9 & 2 & 729.0 & 32.2 & 7.0 \\
        \texttt{Submit} & 7 & 43 & 66 & 7 & 1 & 276.7 & 4.8 & 9.4 \\
        \texttt{TrainPass} & 5 & 14 & 44 & 5 & 1 & 160 & 8.8 & 14.1 \\
        \texttt{Service} & 4 & 8 & - & 4 & 1 & 118.9 & 7.1 & 3.5 \\
        \texttt{FFI} & 3 & 30 & - & 3 & 1 & 73.9 & 6.5 & 1.75 \\
        \bottomrule
    \end{tabular}
    \label{tab:hard}
\end{table}

\subsection{RQ3: Can \polyverl verify multi-language systems?} \label{subsec:rq3}
We show that \polyverl is able to verify multi-language LF programs, ROS2 applications, and programs with inter-language function calls via a foreign function interface (FFI).
Note that these benchmarks cannot be solved by \cite{emsoft23LFVerifier} as they involve multiple languages.
Some results are shown in Table~\ref{tab:hard}.

The final set of LF benchmarks consists of the 22 examples in Section~\ref{subsec:rq1} with half of the C reactions replaced by Rust reactions, plus \texttt{Submit} and \texttt{TrainPass} (the running example in Figure~\ref{fig:lf-train}).
As the performance is similar for the 22 examples, we only report results for the two new benchmarks in Table~\ref{tab:hard}.
The LLM-based synthesizer is able to reason about Rust code and library functions in both C and Rust.

Aside from LF, we can also model and verify ROS2 applications using services (the call-and-response model), shown by the \texttt{Service} example.
The publish-subscribe pattern can be modeled as well.
Another general class of polyglot system involves inter-language function calls, e.g., Rust using C via FFI or Java via Java Native Interface (JNI). 
These systems can be modeled by splitting programs into language-specific segments with typed interfaces (\texttt{FFI} example in Table~\ref{tab:hard}).



%% file: related_work.tex
\section{Related Work}

Most approaches to verifying multi-language systems rely on manually translating some parts of the system until everything is in a common language. 
For instance, Cook et al. \cite{DBLP:conf/fmcad/CookDKMPPTW20} translate fragments of assembly code into C to verify the Xen hypervisor, 
and Marin et al.~\cite{Marin2025} define a small imperative language within the Maude modeling language to manually encode LF reactions.
Where it is common to combine a fragment of one language with another, tools may automate translation of that fragment. 
For example, prior work~\cite{emsoft23LFVerifier} on LF verification automatically translates a fragment of C into the verification language, and tools like CBMC~\cite{cbmc2} support small fragments of assembly code. 
Another method is to black-box unanalyzable code and infer summaries from their observed input-output behavior~\cite{DBLP:conf/esop/Penninckx0P15}.
While similar to us, this relies on sufficient examples and lacks the guarantees that fully integrating language-specific verifiers provides.

We employ compositional verification based on Hoare logic~\cite{DBLP:journals/cacm/Hoare69}. This is commonly applied to verification problems that, whilst written in only one language, are too large to be tackled in a single verification query~\cite{DBLP:journals/pacmse/0001K024}. 
Previous work, e.g., \cite{cousot,moy,gordon,seghir}, automates this by learning pre/post-conditions using abstract interpretation and CEGAR~\cite{DBLP:conf/cav/ClarkeGJLV00}.

There are also property-guided approaches to inferring contracts, while we aim to discover a contract that is just good enough to prove a given property. Here, previous work uses syntactic patterns~\cite{denney}, Constrained Horn Clause solvers~\cite{alshnakat}, and even CEGIS~\cite{DBLP:conf/asplos/Solar-LezamaTBSS06,albarghouthi}. Recently, LLMs have been applied to learning specifications, including pre/post-conditions, for code~\cite{enchantingspecifications}.  
The work closest to ours uses CEGIS and LLMs to infer post-conditions, but neither of these approaches addresses the challenge of multi-language verification.

Our approach to compositional verification, relying on learning-based synthesis oracles, implements the paradigm of ``verification by reduction to synthesis'' and the use of learning-driven synthesis advocated by~\cite{seshia-pieee15}.
Our contract learning approach is based on an integration of CEGIS and CEGAR. This integration can be considered a form of oracle-guided synthesis/learning~\cite{DBLP:journals/acta/JhaS17,seshia-pieee15,DBLP:conf/vmcai/PolgreenRS22}, where the verification oracle and model checker act as oracles guiding the synthesis search. 
In hardware, prior work~\cite{DBLP:conf/vmcai/ZhangYFGM20} uses synthesis to generate environment invariants for modular hardware verification. These invariants can be seen as pre-conditions in our context.

%% file: conclusion.tex
\vspace{-0.2em}
\section{Conclusions}
We formulated the polyglot verification problem and proposed \polyverl, a contract-based approach. This approach bootstraps off-the-shelf language-specific verifiers to verify polyglot systems, while side-stepping challenges involved in compiling the entire system to a single homogenous encoding. 
We implemented our approach in a tool that automatically iterates between CEGAR and CEGIS loops to synthesize contracts using an LLM-based synthesizer.
We demonstrate a significant qualitative improvement by verifying full-fledged, multi-language systems that previous work could not handle. 

There are several interesting directions for future work.
One is to apply the \polyverl approach to verifying other kinds of polyglot distributed software systems or robotics software systems. 
We believe our approach is also applicable in cases where the software is not necessarily polyglot to begin with. For example, it would also be interesting to apply this approach to the problem of formally-verified translation of code from one language to another (e.g., from C to Rust), where the translation is performed in stages, and the intermediate state of the translated code involves a mix of languages.



%% file: acknowledgement.tex
\section*{Acknowledgements}
This work was supported in part by the DARPA Provably Correct Design of Adaptive Hybrid Neuro-Symbolic Cyber Physical Systems (ANSR) program award number FA8750-23-C-0080; the National Science Foundation (NSF) award \#CNS-2233769 (Consistency vs. Availability in Cyber-Physical Systems); Nissan and Toyota under the iCyPhy Center, Intel, Berkeley Deep Drive, and the RDI Frontier Fellows Program.

%% file: appendix.tex
\begin{appendices}

\lstset{
basicstyle=\footnotesize\ttfamily,
columns=flexible,
breaklines=true
}

\section{Large Language Model (LLM) Prompting} \label{appendix:llm}
We use chain-of-thought prompting to produce pre/post-condition contracts for procedures.
The prompt is structured into three stages: (1) task description and instruction, (2) procedure behavior reasoning, and (3) contract generation.

In the first stage, the task is introduced and the requirements are specified.
At this point, a Python-based DSL is also provided so that the LLM can employ it in the later contract-generation stage.

\begin{lstlisting}
Your task is to create two Boolean expressions, 
a precondition and a postcondition, for a function 
named '<function_name>'. Instructions below:
1. Precondition should only depend on inputs.
2. Postcondition can depend on inputs and outputs.
3. Express pre/post-conditions using the following
   set of Python DSL.

# Addition operation
def Add(a, b): return a + b
...
\end{lstlisting}

Next, the LLM performs a code reasoning step.
It is instructed to identify: (1) the assumptions the function might make prior to execution, and (2) the guarantees it provides afterward.
When refining an existing contract, the model is further asked to specify (3) additional constraints that distinguish between a given set of positive and negative examples.
In this refinement setting, we also prompt the LLM to explain why the previous contract was invalid and to propose corrective modifications.
We observe that explicitly reasoning about such additional constraints enables the model to refine contracts more effectively in light of the provided examples.

\begin{lstlisting}
Describe what the function
1. may assume before it is executed, and
2. guarantees after it is executed. Also,
3. think about additional constraints that would 
   satisfy (eliminate) positive (negative) examples.
Output a list of the above points and nothing else.

The function has the following inputs and outputs:
Inputs: <inputs>; Outputs: <outputs>
Code below: <code>

Here is a cex of the following pre/postconditions:
Precondition: <precondition>
Postcondition: <postcondition>
Counterexample: <cex>
Write an explanation of why the pre/postconditions
are violated in the form
```json
{
  "explanation": <EXPLANATION>
  "suggestion": <SUGGESTION>
}
```

Lastly, below are some input/output examples.
Make sure the conditions satisfy the positive
examples and dissatisfy the negative examples.
Compare the positive and negative examples and
deduce what makes the negative examples invalid
behaviors of the code.
List constraints that need to be added to avoid
the negative examples.

Positive example 1: ...
...
Negative example 1: ...
...
\end{lstlisting}

After completing the reasoning step, the LLM is instructed to generate the contract in the provided Python DSL, adhering to the specified requirements.
Any \texttt{SUGGESTION} extracted from its earlier analysis is highlighted in the prompt to guide this generation.
A Python template is then supplied for the model to complete, along with an example to serve as a usage reference.
The final pre/post-conditions are extracted by parsing the Python code block returned in this stage.

\begin{lstlisting}
Based on your analysis, complete the two Python
functions PRECONDITION and POSTCONDITION that
returns pre/postconditions for the code. 
Requirements:
1. The function body should be a single return
   statement. Do not use intermediate variables.
2. Link all fields of the output to the inputs
   if possible, even if they are not used.
3. You may abstract the postcondition by ignoring
   details and focusing on the relationship 
   between the inputs and outputs.
...

**NOTE**: <SUGGESTION> Focus on this when generating
pre/postconditions.

Complete the code below using the provided 
Python DSL and put it in a Python code block.
```python
def PRECONDITION(query):
  return # TODO: Fill in using the provided DSL
def POSTCONDITION(query, response):
  return # TODO: Fill in using the provided DSL
```

Below is an example of using the DSL.
```python
def PRECONDITION(in):
  return TrueExpr()
def POSTCONDITION(in, out):
  return And(
    Eq(Select(out, 'count'), Select(in, 'count')),
    Eq(Select(out, 'is_present), TrueExpr()),
  )
```
\end{lstlisting}



\section{State Space Diagram for LF Program Verification}

\begin{figure*}[htbp]
    \centering
    \begin{subfigure}[b]{\linewidth}
        \centering
        \includegraphics[width=0.7\linewidth]{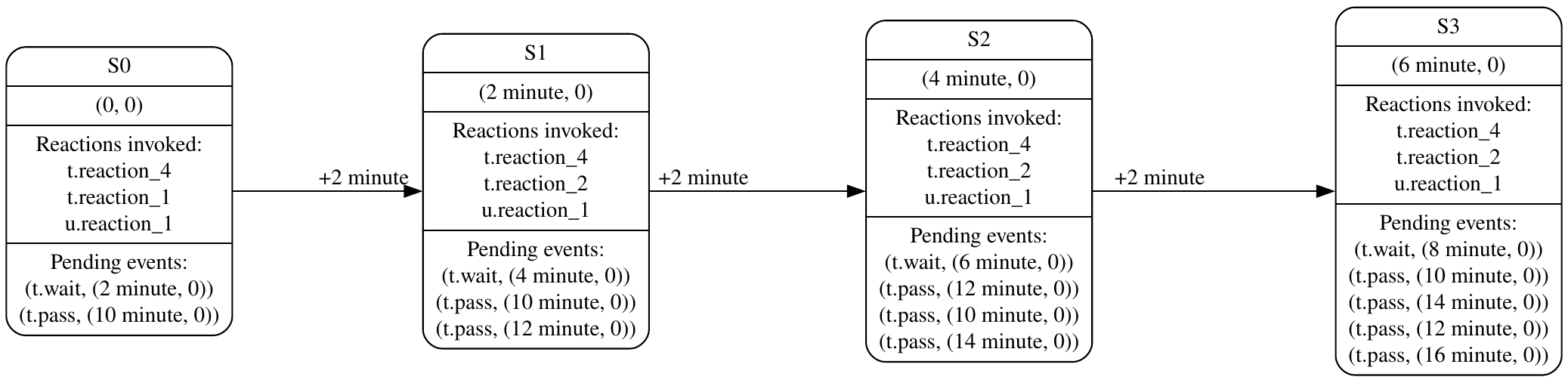}
        \caption{The initialization phase of SSD (including S4)}
        \label{fig:ssd-init}
    \end{subfigure}
    \begin{subfigure}[b]{\linewidth}
        \centering
        \includegraphics[width=0.8\linewidth]{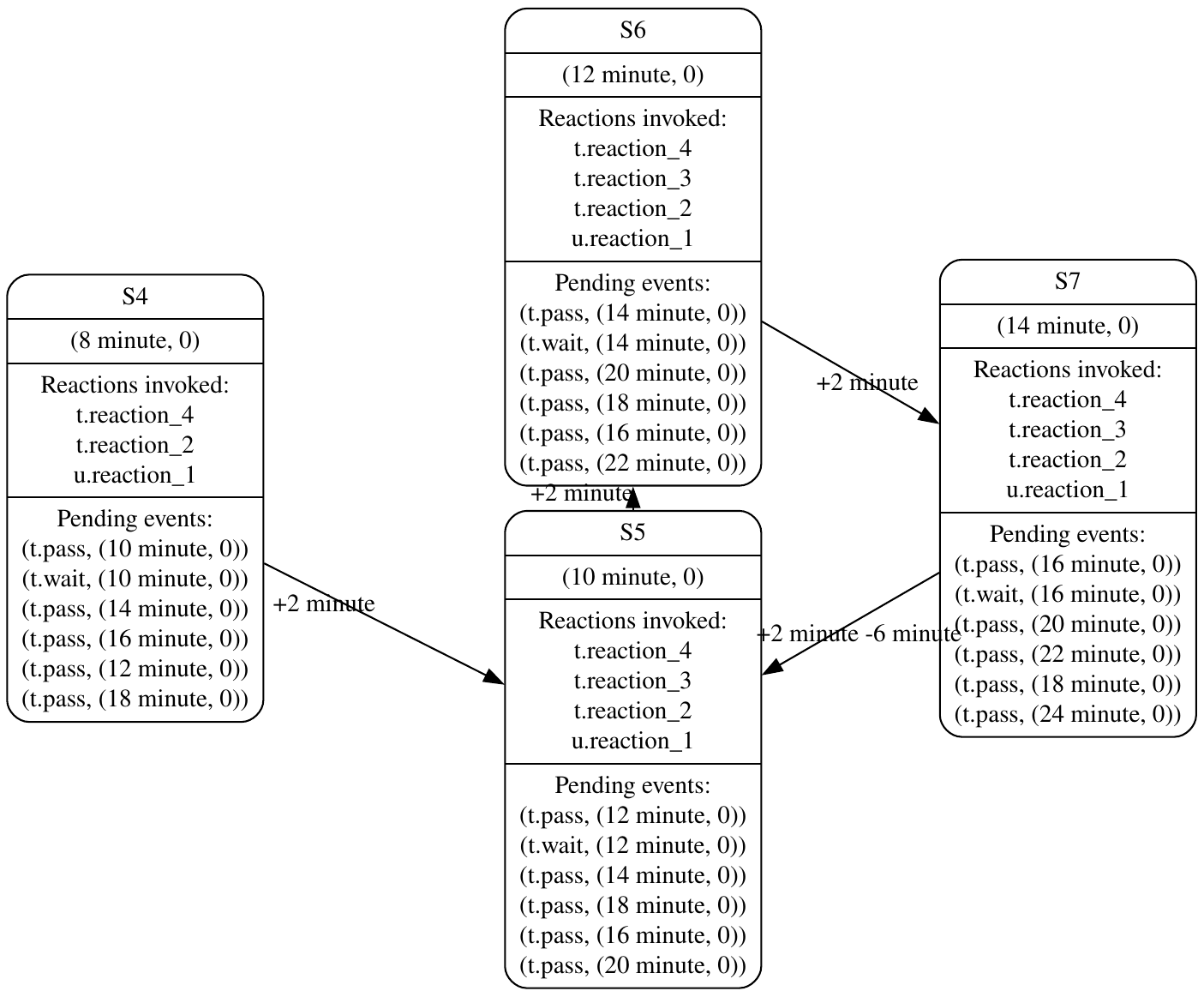}
        \caption{The periodic phase of SSD (excluding S4)}
        \label{fig:ssd-periodic}
    \end{subfigure}
    \caption{The State Space Diagram (SSD) of the LF program in Figure~\ref{fig:lf-train}. The SSD is a lasso, with an initialization phase in Figure~\ref{fig:ssd-init} and a periodic phase in Figure~\ref{fig:ssd-periodic}.
}
    \label{fig:ssd}
\end{figure*}


\end{appendices}